\documentclass[a4paper,fleqn]{cas-dc}

\usepackage[numbers]{natbib}
\usepackage{lipsum}
\usepackage{graphicx}

\begin{document}
\let\WriteBookmarks\relax
\def\floatpagepagefraction{1}
\def\textpagefraction{.001}
\shorttitle{Weak antilocalization in 200-nm HgTe}
\shortauthors{M.L. Savchenko et~al.}

\title [mode = title]{Weak antilocalization in  partially relaxed 200-nm HgTe films}

\author[1,2]{M.\,L.~Savchenko}[orcid=0000-0003-4722-8902]
\cormark[1]

\author[1,2]{D.\,A.~Kozlov}
\author[1,2]{Z.\,D.~Kvon}
\author[1,2]{N.\,N.~Mikhailov}
\author[1]{S.\,A.~Dvoretsky}

\address[1]{Rzhanov Institute of Semiconductor Physics, Novosibirsk 630090, Russia}
\address[2]{Novosibirsk State University, Novosibirsk 630090, Russia}

\cortext[cor1]{Corresponding author. E-mail address: mlsavchenko@isp.nsc.ru}

\begin{abstract}
The anomalous magnetoresistance caused by the weak antilocalization (WAL) effects in 200-nm HgTe films is experimentally studied. 
The film is a high quality 3D topological insulator with much stronger spatial separation of surface states than in previously studied thinner HgTe structures. 
However, in contrast to that films, the system under study is characterized by a reduced partial strain resulting in an almost zero bulk energy gap. 
It has been shown that at all positions of the Fermi level the system exhibits a WAL conductivity correction superimposed on classical parabolic magnetoresistance. 
Since high mobility of carriers, the analysis of the obtained results was performed using a ballistic WAL theory. 
The maximum of the WAL conductivity correction amplitude was found at a Fermi level position near the bulk energy gap indicating to full decoupling of the surface carriers in these conditions. 
The WAL amplitude monotonously decreases when the density of either bulk electrons or holes increases that results from the increasing coupling between surface and bulk carriers.
\end{abstract}


\maketitle

\section{Introduction}

A HgTe-based three-dimensional topological insulator (3D TI) is the ideal platform to probe non-trivial properties  of this class of materials since such structures have leading in the field mobility of two-dimensional (2D) topological surface states, reaching $4\times10^5$\,cm$^2$/Vs~\cite{Kozlov2014}, a bulk energy gap, and a mature technology of creation of gated Hall-bar devices.
Topological surface states in such 3D TIs were widely studied in transport~\cite{Brune2011, Kozlov2014} and magneto-optic responses	~\cite{Dziom2017a, Dantscher2015, Hanson2016}, and by magnetocapacitance spectroscopy~\cite{Kozlov2016}.
It has been established that the surface states in these systems are non-degenerate, and have a non-trivial Barry phase and a quasilinear dispersion.
Moreover, the highest quality of the structures has allowed to study a caused by the topological surface states existence ballistic and interference effects in HgTe-based 3D TIs~\cite{Maier2017, Ziegler2018}.

In the majority of papers devoted to the HgTe-based 3D TIs only 70-100\,nm strained films were studied \cite{Brune2011, Shuvaev2012, Shuvaev2013a, Kozlov2014, Wiedenmann2016, Wiedenmann2017, Maier2017, Ziegler2018}. 
Thanks to uniform strain, these systems have an about 10-15\,meV bulk band gap that makes it possible to separately study the properties of the surface and bulk carriers.
However, the rather short distance between the opposite non-trivial surface states results in number of limitations of checking the response from only one surface of the topological insulator.
Thus, the study of thicker HgTe films is also desirable because of a higher separation between the surface states, their weaker mutual electrostatic coupling, and possible hybridization. 
Recently~\cite{Savchenko2019, Candussio2019}, it has been shown that partially relaxed 200-nm HgTe films have a near zero bulk energy gap, but also host similar to what the strained 80-nm system has -- the spin non-degenerate surface states.

Here we present the study of anomalous magnetoresistance in partially relaxed 200-nm HgTe films at different temperatures and Fermi level positions.
Exploiting the fact of strong sensitivity of the conductivity correction to the coupling between conducting channels (surface and bulk states)~\cite{Kim2011,  Steinberg2011, Chen2011, Kim2013, Lang2013,  Aitani2014, Garate2012, Savchenko2016}, we show that the system under study can be moved to the state of fully decoupled topological surface states.
These results demonstrate the conditions to study properties of spin-polarized electrons from one surface of HgTe-based 3D TIs.

\section{Experimental}

\begin{figure*}
	\centering
	\includegraphics[width=2\columnwidth]{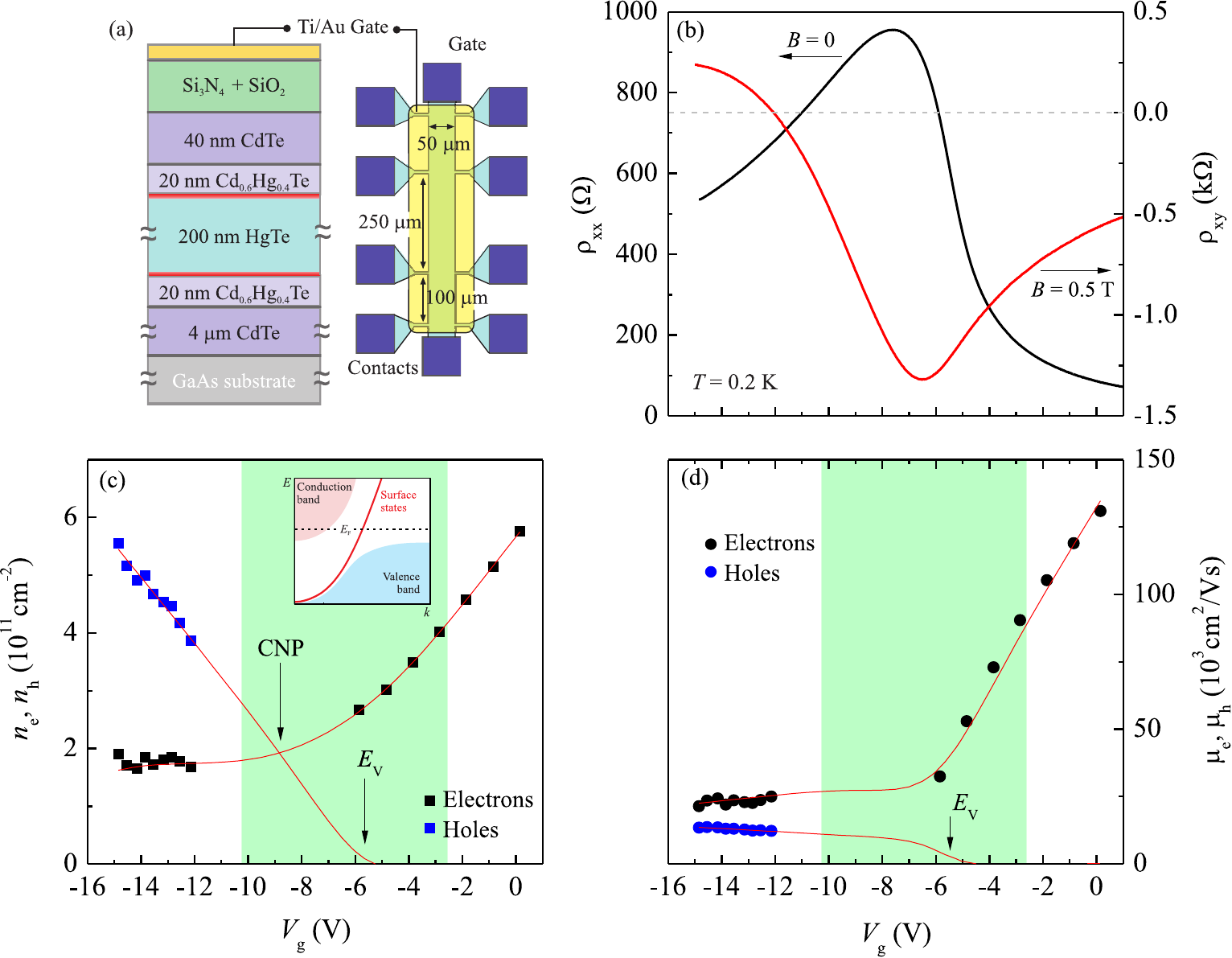}
	\caption{
		(a)~Cross-section of the heterostructure under study, as well as its schematic top view.
		(b)~Gate voltage dependence of $\rho_\text{xx}$ measured at zero magnetic field (black, left axis) and $\rho_\text{xy}$ measured at magnetic field $B = 0.5\,$T (red, right axis).
		(c)~and (d) Gate voltage dependence of obtained from the two-component Drude fitting electron density $n_\text{e}$ and mobility $\mu_\text{e}$ (black squares and circles) and hole density $n_\text{h}$ and mobility $\mu_\text{h}$ (blue squares and circles).
		Charge neutrality point (CNP) and the valence band top $E_\text{v}$ are marked by arrows.
		The inset in panel~(c) shows the schematic band diagram of the system at zero gate voltage.
		The highlighted by green color areas correspond to the gate voltages, at which the
		anomalous magnetoresistance is studied.
		Thin red lines are density and mobility approximations.
	} \label{Fig1}
\end{figure*}

The measurements are carried out on 200-nm HgTe films that have been grown by molecular beam epitaxy on a GaAs~(013) substrate\footnote{
	We note that the same wafer was studied in~\cite{Savchenko2019, Candussio2019}.}.
In Fig.~\ref{Fig1}\,(a) we schematically show cross-section and a schematic top view of the system under study.
The 200-nm HgTe film is sandwiched between thin Cd$_{0.6}$Hg$_{0.4}$Te barriers and is partially relaxed~\cite{Savchenko2019}, a Ti/Au gate has been deposited on the 200-nm Si$_3$N$_4$ plus 100-nm SiO$_2$ insulator grown by a low temperature chemical vapor deposition process.

The schematic band diagram of the partially relaxed 200-nm HgTe film  is presented in inset of Fig.~\ref{Fig1}\,(c).
Previous studies of this structure~\cite{Savchenko2019, Candussio2019} have shown that the system hosts topological states located near top and bottom surfaces of the film (red lines on the cross-section view and in the dispersion) and that by applying gate voltage the Fermi level can be moved from the conduction to the valence band.
The bulk band gap was shown is of the order of 5\,meV.

The studied Hall-bars have a 50-$\mu$m current channel and equal to 100 and 250\,$\mu$m distances between potential probes. 
Transport measurements were performed using a standard low-frequency (about 12\,Hz) lock-in technique with a driving current in the range of 1-50\,nA in a perpendicular magnetic field $B$ up to 3\,T in the temperature range $T = 0.1$-5\,K. 

\section{Experimental results}

\begin{figure}
	\centering
	\includegraphics[width=1\columnwidth]{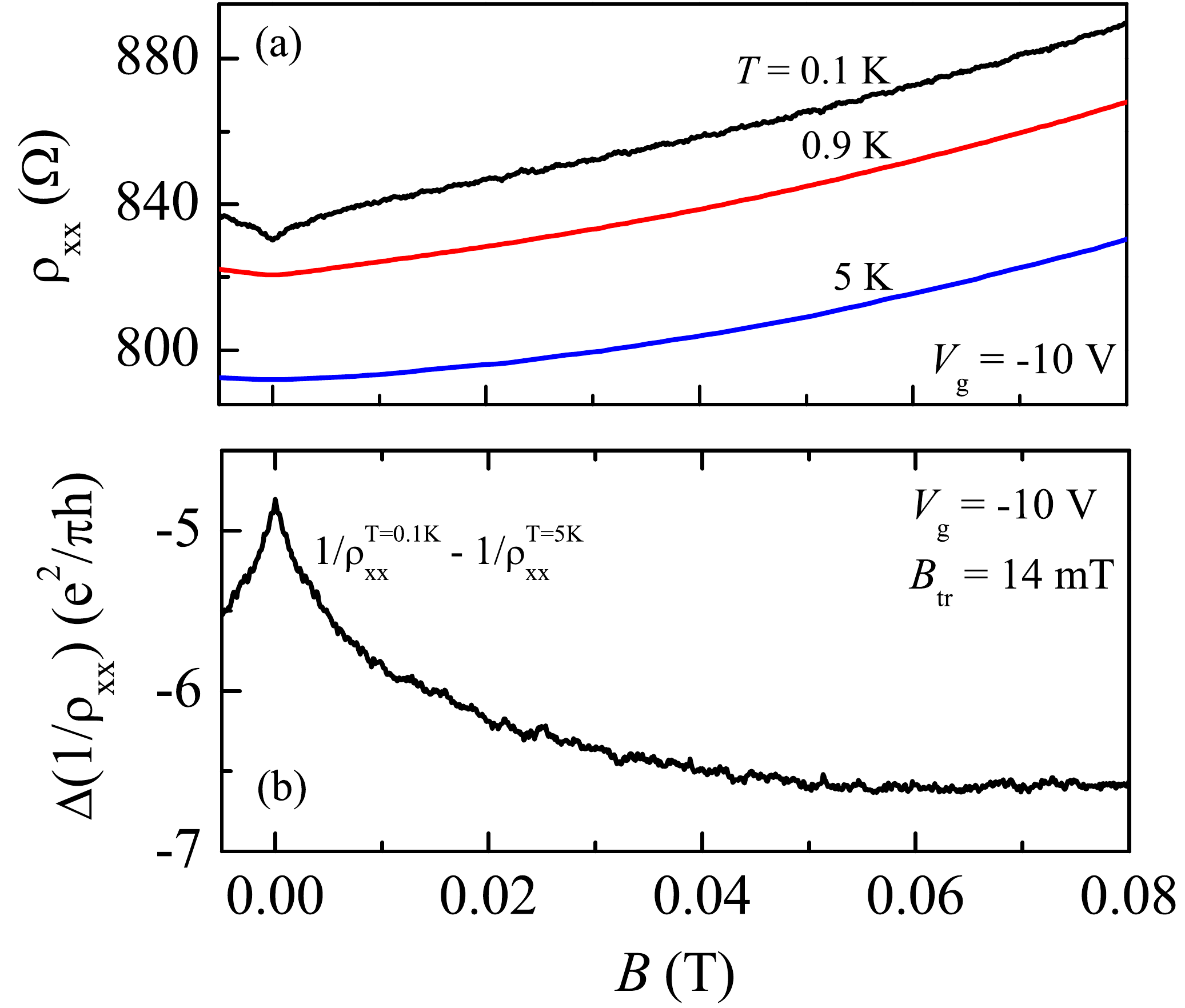}
	\caption{
		(a)~Magnetic field dependence of $\rho_\text{xx}$ measured at $V_\text{g} = -10\,$V and temperatures $T = 0.1, 0.9$ and 5\,K.
		Low-field anomalous magnetoresistance and classic magnetoresistance are seen.
		(b)~Magnetic field dependence of 
		$\Delta (1/\rho_\text{xx}) = 1/\rho_\text{xx}^\text{T=0.1K} - 1/\rho_\text{xx}^\text{T=5K}$ 
		measured at $V_\text{g} = -10\,$V.
	} \label{Fig2}
\end{figure}

In Fig.~\ref{Fig2}\,(b) we present the gate voltage dependence of diagonal resistance $\rho_\text{xx}$ measured at zero magnetic field (black, left axis) and Hall resistance $\rho_\text{xy}$ measured at magnetic field $B = 0.5\,$T (red, right axis).
The $\rho_\text{xx}(V_\text{g})$ dependence exhibits a maximum near $V_\text{g}^\text{max} = -7.5$\,V.
Going from zero gate voltage to its negative values both $\rho_\text{xx}(V_\text{g})$ and $|\rho_\text{xy}(V_\text{g})|$ first goes up, indicating the decreasing of the electron density and mobility, and in the vicinity of $V_\text{g}^\text{max}$ the Hall resistance starts to go down, indicating the emerging of the bulk holes.
$\rho_\text{xy}(V_\text{g})$ changes its sign at about -12\,V that confirms the hole existence.

The qualitative analysis of $\rho_\text{xx}(V_\text{g})$ and $\rho_\text{xy}(V_\text{g})$ is confirmed by the obtained from the two-component Drude fitting of the classic magnetoresistance (not shown) densities and mobilities of carriers.
In Fig.\ref{Fig1}\,(c) we show gate voltage dependence of total electron $n_\text{e}$ (black squares) and hole $n_\text{h}$ (blue squares) densities.
The extrapolation of the hole density to zero results in a point $V_\text{g} \approx -5.5\,$V on the gate voltage axis that corresponds to the Fermi level position near the valence band top.
While to the right from this point the electron density goes up linearly with $V_\text{g}$, it saturates at lower gate voltages, when the Fermi level nearly pinned by high hole density of states (the hole effective mass $m_\text{h} \approx 0.3\,$m$_0$ is about ten time larger compare to electron one $m_\text{e} \approx 0.35\,$m$_0$~\cite{Savchenko2019, Candussio2019}, where m$_0$ is the free electron mass).
In Fig.~\ref{Fig1}\,(d) we show the gate voltage dependences of electron $\mu_\text{e}$ (black circles) and hole $\mu_\text{h}$ (blue circles) mobilities.
The electron mobility, following a Fermi level increase, rapidly grows in the conduction band reaching 10$^5\,$cm$^2$/Vs at $n_\text{e} \approx 5 \times 10^5\,$cm$^{-2}$, and stays nearly constant in the valence band.
The hole mobility is equal to about 10$^4\,$cm$^2$/Vs and also shows nearly no gate voltage dependence. 
The highlighted by green color areas correspond to the gate voltages, at which there is a transition from the conduction to the valence band, and at which the anomalous magnetoresistance is studied.
To perform a quantitative analysis of the interference correction to the conductivity the approximations of $n_\text{e}(V_\text{g})$, $n_\text{h}(V_\text{g})$ and $\mu_\text{e}(V_\text{g})$, $\mu_\text{h}(V_\text{g})$ were used that are shown by thin red lines in Fig.\ref{Fig1}\,(c) and (d).

Despite obtained gate voltage dependences of $\rho_\text{xx}$, $n_\text{e}$, $n_\text{h}$, and $\mu_\text{e}$, $\mu_\text{h}$ are very similar to what was observed earlier in 200-nm films from the same wafer~\cite{Savchenko2019}, there is a significant difference in the value of minimum electron density, when the Fermi level is in the valence band.
Previously, this value was about $0.3 \times 10^{11}$cm$^{-2}$, while now it is about 6 times bigger.
Beside it, now the position of charge neutrality point (CNP) is about at $-9\,$V, while previously it was near the valence band top that was at about zero $V_\text{g}$.
These contradictions can arise from much bigger embedded in the structure during the insulator growth positive charge.
Thus, the bottom topological surface of the studied system has comparable (or exceeding in the valence band) to the top one density and correspondingly contributes to a transport response. 

In Fig.~\ref{Fig2}\,(a) we show a magnetic field dependence of $\rho_\text{xx}$ measured at $V_\text{g} = -10\,$V and temperatures $T = 0.1, 0.9$ and 5\,K.
There is classical positive magnetoresistance at all temperatures, 
which is due to the existence of several groups of charge carriers (bulk holes and surface electrons) and is described within the classical Drude model.
At the lowest temperature there is also a sharp low-field cusp, which shape and amplitude indicates its weak antilocalization (WAL) nature.
We note that a small shift of whole $\rho_\text{xx}(B)$ dependences comes mainly from a classical temperature dependence of $\rho_\text{xx}(B=0)$.
From theoretical point of view it is more conventional to analyze the conductivity correction. 
But since the Drude conductivity itself has a dependence on the magnetic field, from experimental point of view it is more useful to discuss a $1/\rho_\text{xx}(B)$ dependence that has a smaller classic contribution.
To separate out the interference correction we subtract from $1/\rho_\text{xx}(B)$ its classic part, which is measured at higher temperatures (when the quantum correction is fully suppresed), and will work with $\Delta (1/\rho_\text{xx}) = 1/\rho_\text{xx}^\text{T=0.1K} - 1/\rho_\text{xx}^\text{T=5K}$.
Recalculated in terms of $e^2/\pi h$ magnetic field dependence of $\Delta (1/\rho_\text{xx})$ measured at $V_\text{g} = -10\,$V is shown in Fig.~\ref{Fig2}\,(b).
Here it is also clearly seen that there is no transition from WAL to WL because of spin-momentum locking of topological surface states~\cite{Kozlov2016}, and because of a strong 
spin-orbit coupling of bulk HgTe carriers~\cite{Chu2008}.

\begin{figure}
	\centering
	\includegraphics[width=1\columnwidth]{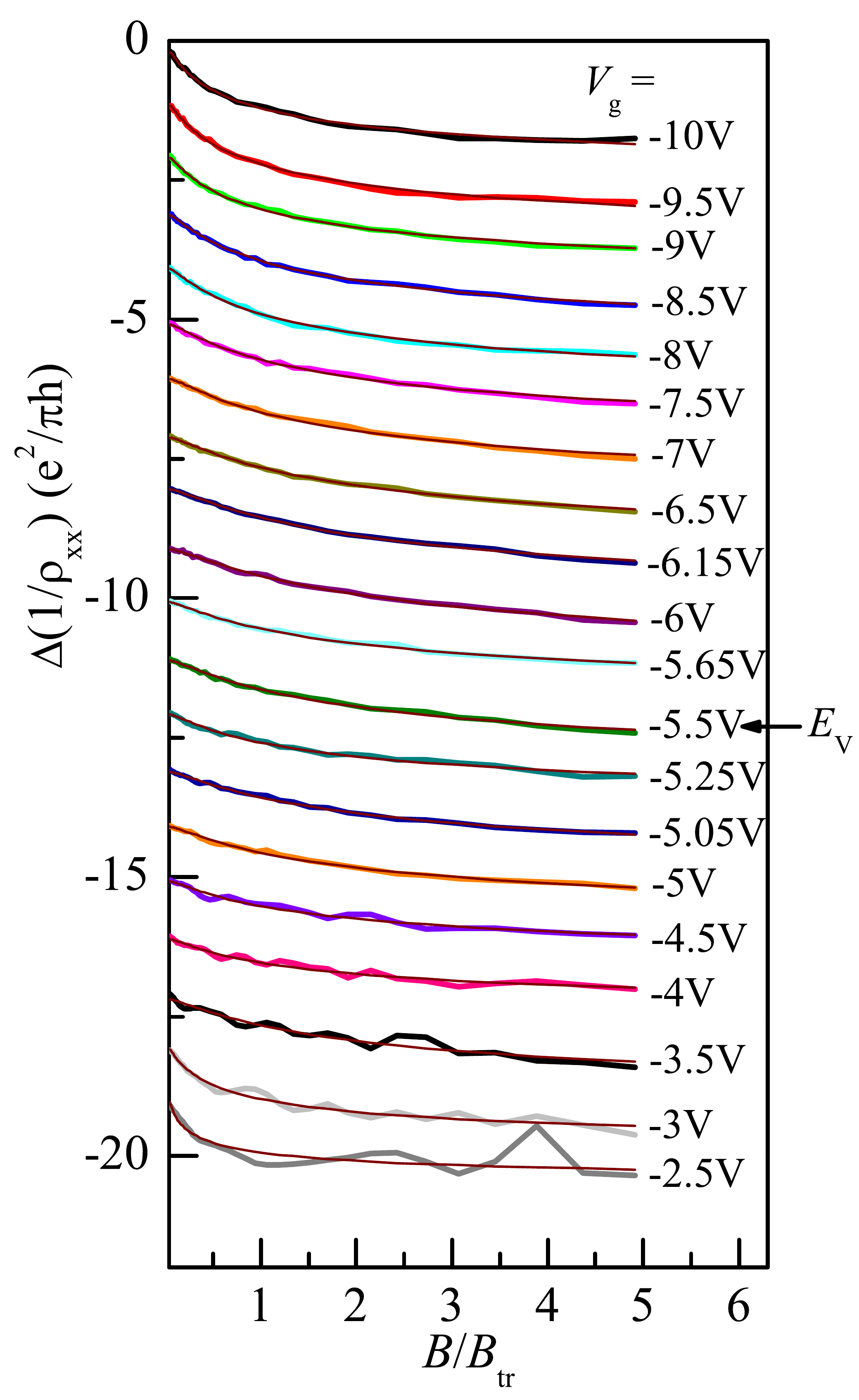}
	\caption{
		Shifted experimental and theoretical~\cite{Golub2005} (smooth) dependences of $\Delta (1/\rho_\text{xx})$ corresponding to		different gate voltages.
		Magnetic field $B$ is normalized by $B_\text{tr}$ for each $V_\text{g}$.
		The arrow indicates the valence band top.
	} \label{Fig3}
\end{figure}

In Fig.~\ref{Fig3} we show measured at different gate voltages $\Delta (1/\rho_\text{xx})$ dependences on normalized magnetic filed $B/B_\text{tr}$ ($B_\text{tr}(V_\text{g})$ dependence is shown in inset of Fig.~\ref{Fig4}\,(a) and will be discussed later)\footnote{
	We note that a close similarity of the presented dependences comes from the magnetic field normalization, since $B_\text{tr}$ has a strong $V_\text{g}$ dependence.}. 
It is seen that at all states of the system the correction to the conductivity is positive as it was in strained 80-nm HgTe films that has about 15\,meV bulk energy gap~\cite{Savchenko2016}.
The absence of a WAL -- WL crossover on the experimental curves is the characteristic feature of 3D topological insulators~\cite{Chen2010, Tian2014, Kim2011}.

\section{Discussion}

\begin{figure*}
	\centering
	\includegraphics[width=2\columnwidth]{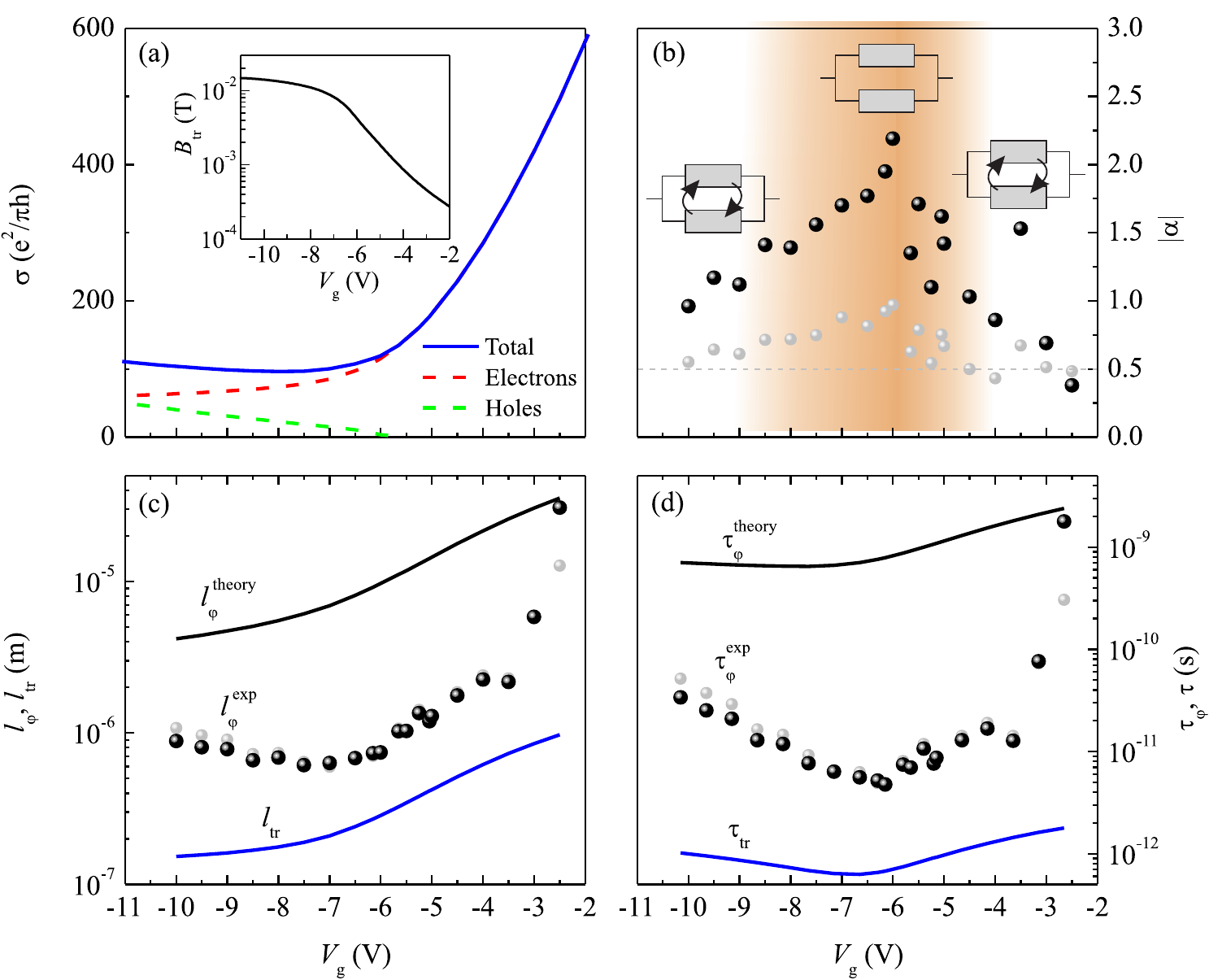}
	\caption{
		(a)~The gate voltage dependence of total (black) and partial electron (red) and hole (blue) conductivity.
		The inset shows the gate voltage dependence of $B_\text{tr} = \hbar/2e l_\text{tr}$.
		(b)~The gate voltage dependence of prefactor $\alpha$.
		Cartoons demonstrate how the existence and absence of the scattering between different types of carriers influences the $\alpha$ value.
		(c)~Experimental $l_\varphi^\text{exp}$ (symbols), calculated $l_\varphi^\text{theory}$ (black line) phase coherence length and transport mean free path $l_\text{tr}$ (blue) versus $V_\text{g}$.
		(d)~Experimental $\tau_\varphi^\text{exp}$ (symbols), calculated $\tau_\varphi^\text{theory}$ (black line) phase coherence time and transport relaxation time $\tau_\text{tr}$ (blue) versus $V_\text{g}$.
		Black symbols on all panels are from the fitting using the ballistic theory~\cite{Golub2005}, gray -- the diffusive Hikami theory~\cite{Hikami1980}.
	} \label{Fig4}
\end{figure*}

The quantitative analysis of the interference correction to the conductivity in 3D topological insulators~\cite{Chen2010, Tian2014, Kim2011, Tkachov2011, Shekhar2014} usually based on the classical Hikami theory~\cite{Hikami1980}. 
Usually, two fitting parameters are in use -- the phase relaxation length $l_\varphi$ (or phase relaxation time $\tau_\varphi$) and the prefactor $\alpha$.
The parameter $\alpha$ is equal to 1 in the ideal case of weak localization and $-0.5$ in the case of weak antilocalization. 
However, when the systems has high mobility carriers, the applicability of the Hikami theory is limited.
Indeed, it works, when $B \ll B_\text{tr}$, where $B_\text{tr} = \hbar/2e l_\text{tr}$ is the transport magnetic field characterizing the boundary between the diffusion and ballistic approximations, and $l_\text{tr}$ is the transport mean free path averaged over all groups of carriers, 
taking into account the weight factors proportional to the contribution of each group $\sigma_\text{i}$ to the total conductivity $\sigma$ of the system~\cite{Savchenko2019,Bykanov1998}.
The gate voltage dependences of partial and total conductivities of the studied system are shown in Fig.~\ref{Fig4}\,(a).
The averaged in such a way $l_\text{tr}$ varies within the range of 0.15-10\,$\mu$m (see Fig.~\ref{Fig4}\,(c)), that results in the $B_\text{tr}$ range from 0.014 to 0.3\,mT (see inset of Fig.~\ref{Fig4}\,(a)).

As it is seen in Fig.~\ref{Fig3}, $\Delta (1/\rho_\text{xx})$ significantly changes near $B/B_\text{tr} \approx 1$ and at higher fields.
Thus, the analysis of $\Delta (1/\rho_\text{xx}) (B/B_\text{tr})$ dependences should be done using a more general theory applicable in both the diffusion ($B/B_\text{tr} \ll 1$) and ballistic ($B/B_\text{tr} \gtrsim 1$) approximations (e.g.,~\cite{Zduniak1997, Golub2005, Glazov2009, Nestoklon2014}).
Despite the theory~\cite{Zduniak1997} was already used in~\cite{Studenikin2003,Studenikin2003a,Savchenko2016}, it does not include all terms in the interference correction~\cite{Averkiev2019}.
Thereby we use a developed in~\cite{Golub2005, Glazov2009} a ballistic theory.
Because of the absence of negative magnetoresistance in the experiment (i.e., there is no WL), the infinite strength of spin-orbit interaction was assumed.
In this case and within the diffusion regime of magnetic fields, the results of this ballistic theory coincides with the Hikami formula, if equal to -0.5 a prefactor value is assumed.

The fitting of the experimental data using the derived in~\cite{Golub2005, Glazov2009} expression\footnote{
	We note that we introduce factor ``-2'' in the formula by hand to deal with the same by meaning ``$\alpha$'' as it is in the Hikami formula.} 
for $\Delta (1/\rho_\text{xx})  = -2\alpha \times \frac{e^2}{\pi h} F \left( \frac{B}{B_\text{tr}}, \frac{l_\varphi}{l_\text{tr}} \right)$ is shown by smooth brown lines in Fig.~\ref{Fig3}.
Since the ballistic WAL correction expression is rather complicated, we first prepared the array of $\Delta (1/\rho_\text{xx}) (B/B_\text{tr})$ dependences at fixed values of $(B/B_\text{tr})$ and $(l_\varphi/l_\text{tr})$.
Then, we fitted the experimental data by this array to find the best suited values of prefactor $\alpha$ and $l_\varphi$.
To calculate the theoretical curves we also used useful notes from~\cite{Nestoklon2014}.

The fitting results in terms of the gate voltage dependences of prefactor $|\alpha|$ and the phase relaxation length $l_\varphi$ (as well as  $\tau_\varphi (V_\text{g})$) are presented in Fig.~\ref{Fig4} by black spheres.
We begin the analysis from the $|\alpha|(V_\text{g})$ dependence.
It is well known that in the simplest systems with strong spin-orbit coupling the prefactor $\alpha$ is equal to $-0.5$.
But here we observe a non-monotonic dependence of $|\alpha|$.
Starting from zero gate voltage, where $|\alpha| \approx 0.5$, it increases, reaching the value of about 1.5	-2 at $V_\text{g} \approx -6\,$V that is close to the valence band top (and, hence, to the bulk gap~\cite{Savchenko2019}).
Going to more negative gate voltages, $|\alpha|$ monotonously decreases and finishes at about 1 at $V_\text{g} = -10\,$V.

The reasonable explanation of the observed $|\alpha|(V_\text{g})$ dependence is based on the Fermi level dependence of the strength of an interaction between different types of carriers in the system. 
It is known that apart from the simplest systems with strong spin-orbit coupling, the prefactor value in a multicomponent structure, such as a 3D topological insulator, 
depends, among other things, on the rate of scattering between different conducting channels~\cite{Kim2011,  Steinberg2011, Chen2011, Kim2013, Lang2013,  Aitani2014, Garate2012}.
In the simplest case, the top and bottom topological surfaces and bulk 
(since the studied system has the close to zero bulk band gap, we suppose there are always either bulk electrons or bulk holes)
can be treated as 2D electron gases with a strong spin-orbit coupling that results in $\alpha_\text{i} = -0.5$ for each group of carriers.
Then, the total expected value of $\alpha$ depends on the scattering rate between the conduction channels: 
if the coupling is weak (or when the time of interchannel scattering is much larger compare to $\tau_\varphi$), the corrections are additive and the expected prefactor value is 
$\alpha = \sum \alpha_\text{i} = -1.5$.
In the opposite case of strongly coupled states they can be treated as single electron gas with $\alpha = \sum \alpha_\text{i}/3 = -0.5$.
Thus, at gate voltages  near to $V_\text{g} \approx -6\,$V, when $|\alpha|$ takes its maximum values, the surfaces and bulk carriers are fully decoupled, and a gradual decrease of $|\alpha|$ to the left and right from this region results in the increase of interchannel scattering. 

The situation, when the topological surface states can be couples through the conducting bulk, is typical for the 3D topological insulators~\cite{Chen2011, Kim2013}.
But at first it seems that the presented results are in opposite to what was observed in strained 80-nm HgTe films~\cite{Savchenko2019}. 
There $|\alpha|$ had minimum values when there were no bulk carriers, and increased, when they emerged.
The first difference can be explained by more than two times bigger space separation of the topological surface states in the 200-nm film compare to its thinner partner.
It results in the strong decreasing of their mutual scattering.
The increase of the surface to bulk coupling in the 200-nm film can be explain by 
the formation of near surface bend bending that modifies the wave functions of the bulk carriers that increases the probability of their scattering with the surface electrons.
When the Fermi level is in the conduction band, the whole system acts as a single 2D electron gas that results in $|\alpha|$ = 0.5.
However, when the Fermi level is in the valence band, localized in band bending bulk holes can effectively scatter only with topological electrons from the top surface. 
Therefore, in these conditions there are two separate conductivity channels -- the bottom topological surface, and top one plus bulk holes.
This results in equal to 1 the absolute value of prefactor.

The fitting results of the experimental curves by the diffusive Hikami theory are shown by gray spheres in Fig.~\ref{Fig4}.
The $|\alpha|(V_\text{g})$ dependence is found to be qualitatively the same compare to the results from the ballistic theory. 
From it one may also conclude a weak scattering rate between topological electrons from top and bottom surfaces, which become strongly coupled with the presence of bulk carriers.
The difference in the values of $|\alpha|$ between this and the ballistic theories comes from too high for the Hikami theory values of $(B/B_\text{tr})$, when this theory overestimates the amplitude of the conductivity correction.

The gate voltage dependence of phase coherence length $l_\varphi^\text{exp}$ and time $\tau_\varphi^\text{exp}$ obtained in experiment are presented in Fig.~\ref{Fig4}\,(c) and (d), respectively. 
In these figures, we also show the theoretical dependences of $l_\varphi^\text{theory}$ and $\tau_\varphi^\text{theory}$ assuming that e-e interaction determines the phase  breaking time $1/\tau_{\varphi}^\text{theory} \simeq
\frac{k_B\,T}{\hbar}\frac{e^2/h}{\sigma}\ln\left(\frac{\sigma}{e^2/h}\right)$~\cite{Minkov2004a}, where $k_\text{B}$ is the Boltzmann constant.
The experimental values lies approximately in the middle between $l_\varphi^\text{theory}$ and $l_\text{tr}$, and between $\tau_\varphi^\text{theory}$ and $\tau_\text{tr}$.
Such a discrepancy is often observed in experiments~\cite{Minkov2012a, Studenikin2003a, Glazov2009, Savchenko2019}
and can be related to a more complex dependence of $\tau_\varphi$ on $\sigma$ in the case of non-parabolic dispersion of carriers.

\section{Conclusion}

To summarize, the anomalous magnetoresistance of the partially relaxed 200-nm HgTe film has been experimentally studied. 
It has been found that the magnetoresistance is positive for all positions of the Fermi level and corresponds to the effect of weak antilocalization. 
It has been shown that the behavior of the anomalous magnetoresistance is well described by the ballistic theory~\cite{Golub2005}. 
It has been found that maximum values of prefactor $|\alpha|$ are observed when the Fermi level is near the valence band top, and there is minimal bulk density~\cite{Savchenko2019}.
In this case $|\alpha| \approx$~1.5-2 that can associated with the absence of mutual scattering between the topological states from top and bottom surfaces and  between surface and bulk conductivity channels.
The filling of the bulk bands is accompanied by a decrease in a prefactor value that indicates the gradual strengthening of scattering between surface and bulk carriers.

\section*{Acknowledgements}

This work supported by RFBR Grant No. 18-42-543013 (together with the Government of the Novosibirsk Region of the Russian Federation). 

%


\bibliographystyle{cas-model2-names}

\bibliography{library}

\end{document}